\documentstyle[pre,aps,epsf]{revtex}
\tighten
\begin{document}
\twocolumn

\wideabs{
\title{Development and geometry of isotropic
and directional shrinkage crack patterns}

\author{Kelly A. Shorlin\cite{kelly} and John R. de Bruyn}
\address{Department of Physics and Physical Oceanography\\
Memorial University of Newfoundland\\
St. John's, Newfoundland, Canada A1B 3X7}

\author{Malcolm Graham and Stephen W. Morris}
\address{Department of Physics \\
University of Toronto\\
60 St. George St., Toronto, Ontario, Canada M5S 1A7}
\date{\today}
\maketitle

\begin{abstract}

We have studied shrinkage crack patterns which form when a thin layer
of an alumina/water slurry dries. Both isotropic and directional
drying were studied. The dynamics of the pattern formation process and
the geometric properties of the isotropic crack patterns are similar
to what is expected from recent models, assuming weak disorder. There
is some evidence for a gradual increase in disorder as the drying
layer become thinner, but no sudden transition, in contrast to what
has been seen in previous experiments. The morphology of the crack
patterns is influenced by drying gradients and front propagation
effects, with sharp gradients having a strong orienting and ordering
effect.

\end{abstract}

}

\section{Introduction}
\label{introduction}

Shrinkage crack patterns are common in both natural and man-made
systems\cite{walker86}, including dried mud layers and the glaze on a
ceramic mug. They are in many instances undesirable, as in the case of
paint or other protective coatings. Crack patterns have long been of
interest in geology in connection with the formation of columnar joints
\cite{basalt,muller98} and ancient crack patterns preserved in the
geological record \cite{k17,t65}. They arise in materials which
contract on cooling or drying. This contraction, coupled with adhesion
to a substrate, leads to a buildup of stress in the material, and when
the stress exceeds the local tensile strength, the material fractures.
This relieves the stress locally along the sides of the crack, but
concentrates stress at the crack tip. As a result the crack tip
propagates until the stress there is reduced below the local strength
of the material \cite{fracture}. A crack pattern forms as multiple
cracks grow and interconnect. In a homogeneous medium a crack will form
and grow perpendicular to the direction of maximum stress. Since the
stress near an existing crack face is parallel to the face, any crack
nucleating at or growing to meet the edge of a preexisting crack will
meet that crack perpendicularly. Non-perpendicular junctions can form
due to the influence of nearby cracks on the local stress field, by the
nucleation of multiple cracks at a point, or by the splitting of a crack
tip into multiple cracks. The lateral length scales of crack patterns
range from millimeters in thin layers of ceramic glaze to tens of meters
in the case of ice-wedge polygons found in the Arctic \cite{l62}.

Shrinkage results from the removal of a diffusing quantity --- water in
the case of dessication, or heat in the case of cooling. In thin layers,
crack development can be slow enough compared to diffusion that the
layer is homogeneous over its thickness and the resulting crack pattern
is effectively two-dimensional. In contrast, in the drying or cooling of
thick layers, gradients of the diffusing quantity --- which can only be
removed from exterior surfaces --- become important. Cracks tend to
propagate in the direction of the gradient, following it as it moves
through the material. This leads to three-dimensional crack patterns.
The crack pattern in turn affects the rate of removal of the diffusing
quantity. Remarkably, an initially disordered network can become
more ordered as it propagates into the material. This leads,
for example, to the formation of hexagonal columnar joints in cooling
basaltic lavas \cite{basalt,muller98}. This ordering process is
currently poorly understood \cite{basalt,muller98}.

In this paper we report on experiments on crack patterns formed by the
drying of slurries of 130 \AA\ alumina (Al$_2$O$_3$) particles
\cite{alumina} in water. We study two-dimensional patterns formed in a
uniformly dried thin layer. A typical crack pattern formed in one of
these experiments in shown in Fig. \ref{layer1}. We also performed
experiments on ``directional drying'' \cite{al95}, in which a
one-dimensional pattern of cracks propagates due to a moving drying
front. A pattern formed by directional drying is shown in
Fig. \ref{directional}.

Although qualitative observations of crack patterns were reported
several decades ago \cite{k17,t65}, there have been rather few
well-controlled experimental studies of the phenomenon. Skejltorp and
Meakin \cite{sm88} investigated the drying of monolayers of
close-packed polystyrene beads. Cracks formed at grain boundaries and
propagated in straight lines defined by the hexagonal lattice of the
beads. Later cracks were wavy. Groisman and Kaplan \cite{gk94} studied
crack patterns in layers of coffee-water mixtures. They found that the
length scale of the pattern was proportional to the layer thickness, $d$,
and increased when the bottom friction was reduced. For thick layers
they observed a polygonal pattern consisting mostly of straight cracks
meeting at 90$^\circ$ junctions. As $d$ decreased, they observed a
transition below which a large fraction of the junctions were at
120$^\circ$ and wavy cracks appeared. A similar transition was observed
in drying corn flour-water mixtures by Webb and Beddoe \cite{wb}.
Pauchard et al. \cite{ppa99} studied the dessication of colloidal silica
sols; in this case the fluid properties are complex and the nature of
the crack patterns was a function of the ionic strength of the
suspension. Korneta et al. \cite{kmm98} studied the geometrical
properties of shrinkage crack patterns formed by sudden thermal
quenches.

There have been several theoretical and computational studies of
two-dimensional crack patterns using spring-block models
\cite{sm88,a94,abj94,la97,msb93,smb94,hsb96,hsb97,cb97,k99}. Andersen
and coworkers \cite{a94,abj94,la97} observed a transition in the crack
pattern morphology as the strain and the coupling to the substrate were
varied. Hornig, Sokolov and Blumen\cite{hsb96}, in a similar model,
observed a change in the way in which fragmentation occurred as the
amount of disorder in the system was varied: for small disorder,
fragments (i.e., polygons) formed through the propagation of cracks in
straight lines defined by the lattice structure. Cracks tended to appear
in the middle of existing segments, so the length scale of the pattern
decreases by a factor of two with each generation of cracks. For
somewhat larger disorder, cracks propagated along wavy paths. For large
disorder, cracks did not propagate but rather formed by the coalescence
of independent point defects. Crosby and Bradley \cite{cb97} found
similar regimes in their simulations, but as a function of the applied
stress due to a sudden temperature quench: low stress led to polygonal
patterns and high stress to a complicated pattern of strongly
interacting cracks. Recently Kitsunezaki \cite{k99} has done theoretical
studies of crack pattern formation, along with numerical simulations
using a random two-dimensional lattice, and reproduced many of the
qualitative features seen in experiments. Kitsunezaki found analytically
that the pattern length scale should be proportional to layer thickness,
as observed experimentally, if the fracture occurs due to a critical
stress criterion, but would be nonlinear if fracture obeyed the
Griffiths criterion \cite{griffiths}.

There have been fewer studies of directional cracking in the presence of
moving gradients in two or three dimensions. Yuse and Sano
\cite{ys93} examined the morphology of {\it single} cracks which
formed in thin glass plates with a moving thermal gradient and observed
an instability of the crack tip as the speed of the cooling front was
increased \cite{ssn94}. Fracture patterns in thin layers of a
directionally dried colloidal suspension between two glass plates have
been studied by Allain and Limat \cite{al95}. Komatsu and Sasa
\cite{ks97} explained the observed regular crack spacing by again noting
that cracks will form in the middle of existing segments where the
stress is highest. Their theory \cite{ks97} predicted a wavelength of
the crack pattern that varied as $d^{2/3}$, in good agreement with the
experimental results of Ref. \cite{al95}, but different from the linear
relation found for the isotropic case.

Recently M\"uller has studied three-dimensional shrinkage crack
patterns in drying cornstarch-water mixtures \cite{muller98}. The
resulting ``starch columns'' are strikingly similar to basalt columns
formed by the cooling of basaltic lava flows. M\"uller showed that the
in-plane length-scale of the pattern (i.e., the column width) was
dependent on $dc/dz$, where $c$ is the water concentration at the
propagating crack front and $z$ is the coordinate along the
propagation direction. From this he inferred that the column width
in basalt is similarly determined by $dT/dz$, where $T$ is the rock
temperature at the crack front.

There has been relatively little theoretical work on three-dimensional
gradient problems.  Motivated by the experiments of Yuse and Sano
\cite{ys93}, Hayakawa\cite{hayakawa} considered two- and
three-dimensional breakable spring models of directional cracking. In
three dimensions, he found columnar structures reminiscent of those
found in basalt\cite{basalt,muller98} and evidence of scaling
behavior in the crack spacing. He did not include adhesion to a
substrate, however.

The remainder of this paper is organized as follows. In  
Section~\ref{experiment}, we describe our experimental apparatus.  In  
Section~\ref{results}, we discuss our results, first on the temporal  
development of uniform and directionally dried crack patterns, and then on  
the geometric characteristics of the final state.  Sections~\ref{discussion}  
and \ref{conclusion} contain a discussion and a brief conclusion.

\section{Experiment}
\label{experiment}

Shrinkage crack patterns were produced by drying layers of a slurry of
130\AA\ Al$_2$O$_3$ particles in water. These particles were very
uniform and inspection of the dried layers indicated that the
composition of the layers was quite homogeneous over their depth.  Two
types of experiments were performed: ``isotropic drying,'' in which
the entire layer was dried uniformly, and ``directional drying,'' in
which the layer was dried from one end.

The isotropic drying experiments were performed by pouring a quantity
of the Al$_2$O$_3$-water mixture into a Plexiglas pan 62 cm
square. The pan was housed in an insulated enclosure. Four 15 Watt
light bulbs mounted 1.6 m above the pan served as a heat source to dry
the films, and also provided illumination for video imaging. The
temperature of the layer was between 25 $^\circ$C and 28 $^\circ$C for
all runs, and was constant to within 1 $^\circ$C for each
run. Time-lapse video recordings of the cracking process were made
using a charge coupled device video camera mounted near the top of the
enclosure. Frames from the video record were digitized for later
analysis, and photographs of the dried layer taken at the end of each
run were also analyzed. Experiments were performed with with the
Plexiglas substrate untreated, as well as with the substrate coated with a
thin transparent coating of teflon to reduce the friction between the
drying layer and the substrate. Experiments were also done with the
teflon-coated surface and ``impurities,'' in the form of 10 cm$^3$ of
sand grains, 0.425--0.500 mm in size, sprinkled evenly over the top of
the slurry before the drying began. The sand floated on top of the
layers during drying. These impurity particles were $\sim 3 \times
10^3$ times the size of the alumina particles and of the same order as
the layer thickness. The analysis discussed in Section \ref{results}
was done on 8 cm square sample areas within the pan away from the
edges.

Directional drying experiments were performed on layers of
Al$_2$O$_3$-water mixture placed in a 12 cm $\times$ 8 cm rectangular
pan. The pan was covered with a sealed window, leaving a 3 mm air
space above the layer. A very slow laminar breeze of ultradry N$_2$
gas, 20$^\circ$ above ambient temperature, was passed unidirectionally
through the air space. The N$_2$ became saturated with water as it
passed through the cell, resulting in a drying front that moved
downwind.  By controlling the rate of flow, the front speed could be
varied between 0.01 mm/hr and 10 mm/hr. The progress of the front was
monitored with time-lapse video.  The roughness of the substrate vas
varied by putting teflon and various grades of sandpaper on the bottom
of the pan. Our apparatus differs from that of Allain and
Limat\cite{al95} in that here the upper surface of the drying layer is
free and only the lower surface is constrained by adhesion to the
substrate.  We found that a considerable transport of material
occurred over the course of an experiment, so that the final dry layer
could be as much as 50\% thicker at the downwind end than at the
upwind end. This could be compensated for by tilting the apparatus.

\section{Results}
\label{results}

\subsection{Dynamics}
\label{Pattern development}

The alumina-water mixture is initially fluid, but becomes gel-like a
few hours after being poured into the experimental container. Cracks
first appear in the layer after a wait of from several hours to a few
days, depending on the thickness and composition of the layer. The
drying process in the ``isotropic" experiments is, of course, neither
perfectly isotropic nor perfectly uniform as a result of
nonuniformities in thickness and in the heating of the layer by the
light bulbs. Because of this, cracks tend to start in one or more
separate regions of the layer and move into the rest of the system, so
that different parts of the system are at different stages of
crack-pattern evolution at any given time. The formation of the crack
pattern and complete drying of the layer took up to eight days, again
varying with layer thickness.

The dynamics of crack pattern formation were observed on time-lapse
video recordings. Although some characteristic behavior was observed, in
some cases paralleling predictions of the theoretical models discussed
above, there was no single clear-cut path followed during the
development of the patterns.

In the first stage of crack pattern development, cracks nucleate at a
small number of points in the layer. In the absence of sand
impurities, these cracks then propagate in fairly straight lines,
presumably in a direction determined by the local drying
conditions. In runs with sand added to the layer, cracks tend to
nucleate at the sand grains and in many cases do not propagate long
distances across the layer, but rather move in short steps from one
sand grain to the next, occasionally meeting another crack moving in a
similar fashion. In the runs with no sand, other defects in the layer
structure presumably serve as nucleation sites. Typically, symmetric
arrangements of two or three cracks nucleate at a site and propagate
away from it. Images taken from time lapse video recordings showing a
number of triplets of
cracks in the early and later stages of the
development of a crack pattern are shown in Fig. \ref{triplet}. More
examples can be found in Fig. \ref{layer2}, which shows a pattern from
a run in a thin layer with sand impurities. In runs with no sand
impurities, the number of these nucleations is small. After the
initial nucleation of a few cracks, most new cracks start at the sides
of existing cracks and propagate away from their parent crack at right
angles.

The longest cracks in Fig. \ref{layer1} are an order of magnitude longer
than the final length scale of the pattern. These long, straight cracks
are referred to as primary cracks. Although these may not be the very
first cracks to form in the layer, they are usually the oldest cracks
locally. As the slurry dries further, successive generations of cracks
form, often joining older cracks and forming a  complicated array of
polygons. The resulting pattern has a characteristic length scale, as
can be seen in Fig. \ref{layer1}. The polygonal fragments bounded by the
cracks are predominantly four-sided. Most of the junctions between
cracks are at right angles, with a few non-perpendicular junctions. In
runs with no sand impurities, the faces of the cracks are smooth and the
cracks, while they can curve smoothly, are usually locally straight.

In thinner layers, or layers with sand impurities added, the patterns
are more disordered in appearance. An example is shown in Fig.
\ref{layer2}. A few long primary cracks are still present, but are
much less dominant than in Fig.  \ref{layer1}. In runs with sand
added, cracks can change direction suddenly, and in thin layers the
crack faces appear less smooth. Despite the more disordered
appearance of the pattern in Fig. \ref{layer2}, the fragments of
dried mud are still predominantly four-sided and crack junction angles
remain primarily at 90$^\circ$, as discussed below.

A common process in the isotropic system we call
``arching."  In this process, a new crack propagates away from an
existing pattern of cracks. It then curves back on itself in a roughly
parabolic path, eventually stopping as it runs into another crack. The
region enclosed by the arch is fragmented by later cracks as it dries
further. The stages in this process are illustrated in Fig. \ref {arch}.

Another very commonly observed mode of pattern formation is shown in Fig.
\ref{ladder}; we refer to this process as ``laddering." In this
process, two or more cracks propagate more or less parallel to each
other, following a local drying front. As the layer behind the moving
crack tips dries further, perpendicular cracks form, joining two of
the parallel cracks like the rungs of a ladder. Fig. \ref{ladder}
shows the stages in this process, and regions in which extensive
laddering have taken place can be seen in Fig. \ref{layer1}. This
process is dominant in the directional drying case, as can be seen in
Fig. \ref{directional}, and clearly involves a local directionality of
the drying process in an essential way.

Using the directional drying technique, we could examine the
laddering process in a more controlled way. The directional experiments
involved much sharper fronts than those that emerged spontaneously in the
isotropic version of the experiment.
Fronts produced alignment and a uniform spacing similar to that
observed by Allain and Limat\cite{al95}, although on a larger
scale.  By varying the rate of drying, we found that the spacing of
the primary cracks was independent of the speed of front advance
over 3 orders of magnitude.  Once the spacing of the primary cracks was
established perpendicular to a planar front, we did not observe it to change 
subsequently.  In particular, we do not normally observe the sequential  
bisection proposed in Ref.~\cite{ks97}. The scale of the primary cracks in  
turn determined that of the secondary, ``rung'' cracks of the ladder.

The primary crack spacing in the directional experiments
 does not have a simple relationship
to the local depth, because the propagation of the front introduces a form  
of history-dependence.  We constructed a cell which contained a 0.5~mm  
downward step and examined the 
 primary crack pattern as it passed over the  
step.
The subsequent behavior of the primary crack spacing was strongly  
influenced by the roughness of the substrate. For the teflon-covered  
substrates, no new primary cracks appeared as the front passed the step, as  
illustrated in Fig.~\ref{directional}. For substrates covered with the  
roughest grades of sandpaper, under otherwise identical conditions, the  
primary crack spacing doubled as it passed over the step, with a primary  
crack disappearing
between each of the original ones.  This is shown in
Fig.~\ref{directional_step}.  This demonstrates that the primary crack
spacing is, in a sense, hysteretic on rough substrates, with a range of
spacings which can stably propagate under given conditions.

The theories of Refs. \cite{hsb96} and \cite{ks97} predict that
cracks will tend to form in the middle of existing fragments of the
layer, since that is ideally where the stress in the fragment will be
largest. In our isotropic experiments new cracks are observed to form at almost
any location in an existing fragment.  They usually start at one edge,
rather than in the interior, and propagate more or less in a straight
line across the fragment. It is rare that the new crack will exactly
bisect the fragment, presumably due to nonuniformities in the adhesion
of the fragment to the substrate or in its dryness or
composition. This is evident from the range of sizes of the fragments
seen, for example, in Fig. \ref{layer1}. However on occasion
near-perfect bisections do occur; one example is shown in Fig.
\ref{bifurc}. In this example the regions between propagating parallel
cracks are bisected by new cracks, as proposed for the directional
drying case in Ref. \cite{ks97}.

In the very late stages of crack
pattern formation, new cracks tend to simply cut off the corners of
existing fragments, breaking off a piece much smaller than the
original fragment. The results of this process can be seen in many
places in Fig. \ref{layer1}.

\subsection{Geometrical characteristics}
\label{geom}

The characteristic length scale of the final crack pattern was
determined in real space as well as from a Fourier analysis of the
patterns. We estimate the pattern ``wavelength'' $\lambda_r$ by
$l/\sqrt{N_{p}}$ where $N_{p}$ is the number of polygons in a square
region of the pattern of side $l$. For a perfect pattern of
equal-sized squares, this would be equal to the true
wavelength. $\lambda_r$ was measured in a number of 8 cm by 8 cm
square regions, neglecting the parts of the layer close to the edges,
and is plotted as a function of the mean local layer depth $d$ in
Fig.~\ref{lambda_vs_d}. In all three sets of isotropic drying
experiments (untreated substrate, teflon-coated substrate, coated
substrate plus added impurities) $\lambda_r$ is proportional to depth,
as found previously by Groisman and Kaplan \cite{gk94}. The slope is a
function of the experimental conditions, with the runs having lower
friction between the layer and the substrate
(Fig. \ref{lambda_vs_d}(b)) having the highest slope. Both increased
friction (Fig. \ref{lambda_vs_d}(a)) and the addition of impurities
(Fig. \ref{lambda_vs_d}(c)) cause the slope to decrease. Equivalently,
at a given layer depth, increased friction and the addition of
impurities lead to a decrease in the length scale of the pattern. The
scatter in the data increases for depths greater than about 2.75 mm,
particularly for the runs with reduced friction and no
impurities. This may be a result of the decreased effect of bottom
friction with thickening layers, or of three-dimensional effects.

The distribution of length scales in the pattern was also studied by
performing two-dimensional FFTs of video images of the dried layer.
This analysis does not require any assumptions about the shape or
orientation of the polygons making up the pattern. The two-dimensional
Fourier power spectrum $P_1(k)$ of an image, where $k$ is the
wavenumber, was calculated. A large low-$k$ peak, due to spatial
nonuniformities in the illumination, was removed and the power
spectrum was azimuthally averaged. The resulting spectrum is peaked
about a wavenumber $k_{c}$, and $\lambda_c = 2 \pi/k_{c}$ can be taken
as the characteristic length scale of the crack pattern. The spectra
were asymmetric, with the skewness of the distribution being close to
zero for the thinnest layers studied and increasing with
depth. Fig.~\ref{lambda_vs_d} also shows the values of $\lambda_c$
found in this way.  The two methods of finding the length scale of the
pattern are consistent.

As can be seen in Figs. \ref{layer1} and \ref{layer2}, most polygons
in the final pattern are four-sided, and most crack junction angles
are at 90$^\circ$. Even non-four-sided polygons tend to have
90$^\circ$ corners; this is possible because the cracks 
which define
their sides are curved. Examples of this can also be seen in
Fig. \ref{layer1}. Fig.  \ref{nsides} shows the distribution of
$n$-sided polygons $P_2(n)$ averaged over all depths studied, for runs
with the untreated substrate, the reduced friction substrate, and the
treated substrate with added sand impurities. The distributions were
essentially independent of depth over the range studied ($0.13$ mm
$\le d \le 4.12$ mm), except as noted below, and the distributions in
the three cases are the same within the experimental
uncertainties. With both the treated and untreated substrates (but
without impurities), there is some indication of a drop in the
fraction of three-sided polygons and a corresponding jump in the
fraction of four-sided polygons as the depth increases above 1.5 mm.
If this variation (which, although persistent, is within the
statistical uncertainty of the data) is real, it is probably a result
of cracks breaking off the corner of an existing polygon in the late
stages of pattern formation, as described above. This process can
turn, for example, a four-sided polygon into a three- and a five-sided
polygon. This phenomenon is less common in thicker layers and in runs
with added impurities.

The distribution $P_3(\theta)$ of crack junction angles $\theta$ was
sharply peaked at 90$^\circ$ for all runs. Sample distributions, using
5$^\circ$ bins for $\theta$, are shown in Fig. \ref{angles}. Fig.
\ref{angles}(a) shows $P_3(\theta)$ for runs with the untreated
substrate, for a relatively thin layer and a relatively thick
layer. The distributions are essentially identical. In particular, the
mean is 90$^\circ$ within the experimental uncertainties and the width
of the distribution does not change with depth. Fig. \ref{angles}(b)
shows distributions for runs with the reduced-friction
substrate. Again the mean is $90^\circ$ for these runs, but in this
case the distribution become much broader for thinner layers than for
thicker layers. When sand was added to the layer, the mean remained at
90$^\circ$ and again the distribution tended to broaden as $d$
decreased. None of our runs had clear peaks at angles other than
90$^\circ$; the broadening of the distribution for small $d$ was due
to a general spreading of the distribution and not to the emergence of
peaks at, for example, 120$^\circ$. Although symmetric triplets of
cracks with junction angles of 120$^\circ$ were seen to nucleate in
the early stages of drying, as discussed above, the number of such
events was always too small to show up as a significant peak in the
distribution functions, even in runs with sand added.

\section{Discussion}
\label{discussion}

Cracks develop in a drying layer when the shrinkage of the layer
induces sufficient stress that the layer fractures. Our observations
indicate that cracks initially appear by nucleation at a few
points. When sand impurities were added to the layer, sand grains were
frequent nucleation sites. Triplet junctions, at which three cracks
meet with junction angles of 120$^\circ$, are formed by such
nucleations in the early stages of pattern development. At later
times, however, cracks meet predominantly at 90$^\circ$
junctions. This is due to the fact that cracks propagate in the
direction which most efficiently relieves the stress. Since the stress
near a given crack is parallel to its surface, other cracks will tend
to approach and meet it at right angles.

The formation of crack patterns is strongly influenced by drying
gradients, which in turn lead to stress gradients.  Cracks tend to
propagate in the direction of a dryness gradient. This is most obvious
in the patterns produced in the directional drying experiments, but is
also the cause of the laddering patterns observed in the isotropic
experiments. The arching illustrated in Fig. \ref{arch} presumably
also results from the interacting between the propagating crack and
the local dryness or stress field.

Theories of crack pattern formation \cite{hsb96,ks97} indicate that
each generation of cracks should appear in the middle of existing
fragments, leading to successive halvings of the length scale of the
pattern. This follows from the fact that ideally the stress in a
fragment of the drying layer will be a maximum midway between two
cracks. Near-perfect halving of existing fragments, as in
Fig. \ref{bifurc}, was seen only rarely in our experiments. This is
presumably due to unevenness in the local stress distributions, in
turn resulting from nonuniform adhesion to the substrate and
nonuniformities in the dryness and composition of the layer.

The spacing of the primary cracks in the directional drying experiments
did not exhibit the halving effect either.  Instead, the spacing that was  
quickly established near the beginning of the run propagated stably, as if we  
only observed the endpoint of the halving process.  Even when sudden changes  
in stress were forced by propagating fronts over steps in the depth, spacing  
changes only occurred for rougher substrates.

The length scale of the final crack pattern in the isotropic
experiments was found to be proportional to the layer depth, as shown
in Fig. \ref{lambda_vs_d}, with the constant of proportionality being
largest for reduced friction between the layer and the substrate. This
is in agreement with the experiments of Groisman and Kaplan
\cite{gk94} and also with the theoretical work of Kitsunezaki
\cite{k99} under the assumption that the cracks develop as a result of
a critical stress mechanism. This is easily understood. With reduced
substrate adhesion, the stress in the layer will grow more slowly with
distance away from an existing crack, and so a larger region is needed
for the layer to reach the critical stress for fracture in that
case.


The data in Fig. \ref{nsides} show that the distributions of the
number of sides of the polygons in the final pattern are essentially
the same for the experimental conditions studied. The distribution of
junction angles was sharply peaked at $90^\circ$ in all runs. For
experiments with the untreated substrate, there was no significant
variation in the distribution of angles with depth, as indicated in
Fig. \ref{angles}(a). The standard deviation of the distribution was
about 15$^\circ$ for all depths, and no peaks were observed at any
other angles. With reduced bottom friction, both with and without
added sand, the distribution broadened as the layer depth decreased,
as indicated in Fig. \ref{angles}(b). The standard deviation of the
distribution increased from approximately 5$^\circ$ for layers 2.5 mm
thick to 20$^\circ$ for layers 0.7 mm thick. Although nucleation of
triple cracks was observed in the early stages of pattern formation,
particularly with added sand, there was no sign of significant peaks
in the distribution at angles other than 90$^\circ$.

For comparison, Groisman and Kaplan \cite{gk94} showed a distribution
of junction angles for a thick layers of dried coffee-water mixture
which has a mean of 93.1$^\circ$ and a standard deviation
8.0$^\circ$. Groisman and Kaplan \cite{gk94} observed a transition in
the distribution of junction angles as layer depth was decreased: for
layers thinner than 4 mm, they observed a marked increase in the
number of 120$^\circ$ junctions. In their thinnest layers, they found
up to 30\% 120$^\circ$ junctions. They explained this as being due to
an increased likelihood of nucleation of cracks from inhomogeneities in
the material when the layer thickness became roughly the same as the
size of local inhomogeneities. While we do see a broadening of the
distribution of angles when the substrate is coated with teflon, we
observe no sharp transition down to layer depths of 0.5 mm, which is
the size of the sand grains added to our mixture. We also see no
evidence for a peak in the distribution at angles of $120^\circ$.  It
is likely that our alumina slurries were much more homogeneous than
the coffee-water mixtures used in Ref. \cite{gk94}, but this does not
explain the absence of an increase in the number of triplet junctions
in thin layers with added sand.

The theoretical models of Refs. \cite{hsb96} and \cite{cb97} predict
that the appearance of the crack pattern should change as the amount
of disorder in the system increases. With no sand added, our layers
are very homogeneous and patterns are formed by cracks which
propagate in straight lines and interact to form an array of
predominantly four-sided polygons. These patterns are consistent with
the model predictions in the case of weak disorder \cite{hsb96}. With
sand added, and to some extent in thinner layers, the cracks tended to
propagate shorter distances and the patterns became closer to the type
predicted for medium disorder. Thus our results provide a qualitative
confirmation of the validity of the models.

\section{Conclusions}
\label{conclusion}

Our isotropic drying experiments resulted in crack patterns which were
similar in appearance to those predicted by theoretical models in the
case of weak disorder. This is consistent with our expectation that
the alumina-water mixtures used in our experiments were quite uniform
in composition down to rather small length scales.  The development of
the crack patterns in our experiments did not proceed by a single
well-characterized process.  Initially, cracks nucleate at points.  In
layers with sand added, sand grains often serve as nucleation
sites. The distribution of crack junction angles is strongly peaked at
90$^\circ$ for all experimental conditions studied. 120$^\circ$
triplet junctions can result from the symmetric nucleation of three
cracks at a point, but most crack junctions form later in the
development of the pattern when a propagating crack meets an older
crack at 90$^\circ$. We do not observe a dramatic increase in the
fraction of triplet junctions nucleated as the layer gets thinner, in
contrast to the observations of Ref. \cite{gk94}. Models predict that
new cracks will form midway between existing cracks \cite{hsb96,ks97},
leading to a halving of the pattern length scale with each successive
generation of cracks. Although this process does sometimes occur, it
is not predominant, at least partly because of nonuniformities in
layer thickness, substrate adhesion, and drying. Other processes such
as the laddering and arching described above are much more common.  The  
laddering effect is most pronounced in the directional drying experiments.   
We found that the primary crack spacing in the directional experiments did  
not show the halving effect either, but rather that the spacing tended to  
become established early and then propagate stably thereafter, even over  
sudden changes in depth.  The spacing in the directional experiments depended  
on the substrate adhesion, but not on the speed of front propagation.

In summary, we have performed an extensive experimental study of crack
pattern formation in both the isotropic and directional drying cases.
Our results provide some confirmation of the predictions of theoretical
models, but also underline the complexity of the pattern formation
process in this system.

\section*{Acknowledgements}
\label{ack}
This research was supported by NSERC of Canada. We acknowledge helpful
discussions with N. Rivier, P.-Y. Robin, and development work by
Eamonn McKernan.

\begin{figure}
\caption{ A portion of the crack pattern formed by drying a 62 cm
square layer of an alumina-water mixture. The area shown measures 16.0
cm by 16.0 cm. The Plexiglas substrate was untreated and the final
thickness of the layer was $2.27 \pm 0.20$ mm.  }
\label{layer1}
\end{figure}

\begin{figure}
\caption{The crack pattern formed in a directional drying
experiment. The substrate was treated with teflon and a sharp drying front  
moved front left to right across the layer, leaving behind oriented primary  
cracks. The rungs of the ladder-like pattern are secondary cracks. The front  
speed was 1.2 mm/hr.  Near the middle of the picture, the depth of the layer  
suddenly increased by about 50\%, but no change in the average spacing is  
evident. This illustrates the weak depth dependence and hysteretic nature of  
crack spacings on slippery substrates. }
\label{directional}
\end{figure}

\begin{figure}
\caption{A portion of the crack pattern forming in a layer $1.31 \pm
0.03$ mm thick. Sand has been added to the layer and the substrate was
coated with teflon. At least five triplet junctions, formed when three
cracks nucleated at a point, are visible in the images. The area shown
is 5.7 cm by 5.5 cm. (a) 41 hours into the run. (b) 56 hours
into the run.  }
\label{triplet}
\end{figure}

\begin{figure}
\caption{A portion of the crack pattern formed by drying a 62 cm
square layer of an alumina-water mixture. The area shown measures 17.2
cm by 17.2 cm. The Plexiglas substrate was coated with teflon and sand
was added to the mixture. The final thickness of the layer was $0.60
\pm 0.03$ mm. }
\label{layer2}
\end{figure}

\begin{figure}
\caption{A sequence of images of a portion of a crack pattern forming
in a layer of final depth $1.51 \pm .06$ mm with a teflon coated
substrate.  The field of view is 5.1 cm by 4.8 cm.  Image
(a) is 33 hours after the start of the run and the elapsed time
between (a) and (f) is 6 hours. This sequence illustrates the
``arching'' process described in the text. }
\label{arch}
\end{figure}

\begin{figure}
\caption{A sequence of images of a portion of a crack pattern forming
in a layer of final depth $1.88 \pm .10$ mm, with an uncoated
substrate. The field of view is 7.1 cm square.  Image (a) is 67 hours
after the start of the run and the elapsed time between (a) and (f) is
17 hours. This sequence illustrates the ``laddering'' process
described in the text.}
\label{ladder}
\end{figure}
\noindent 

\begin{figure}
\caption{A directional drying experiment on a rough sandpaper
substrate. The sharp drying front moved from left to right across the
layer at a speed of 1.5 mm/hr.  Near the middle of the picture, the
depth of the layer suddenly increases by about 50\%, as in
Fig.~\protect{\ref{directional}}. At this point, the crack spacing
changes abruptly, approximately doubling.}
\label{directional_step}
\end{figure}
\noindent 

\begin{figure}
\caption{A sequence of images of a portion of a crack pattern forming
in a layer of final depth $2.50 \pm 0.14$ mm with a teflon coated
substrate.  The field of view is 9.7 cm square.  Image
(a) is 135 hours after the start of the run and the elapsed time
between (a) and (c) is 5 hours. This sequence illustrates the
formation of cracks which bisect regions bounded by existing cracks,
as predicted in the theory of Ref. \protect\cite{ks97}.
}
\label{bifurc}
\end{figure}

\begin{figure}
\caption{The pattern wavelength as a function of layer depth for (a)
untreated substrate, no impurities; (b) reduced substrate friction, no
impurities; and (c) reduced substrate friction with added
impurities. Open symbols are values of $\lambda_r$, determined
geometrically, and solid symbols are values of $\lambda_c$, determined
from a Fourier analysis of the pattern.  }
\label{lambda_vs_d}
\end{figure}

\begin{figure}
\caption{The distribution of $n$-sided polygons in the final pattern,
averaged over depth, for the three different types of isotropic drying
runs. Error bars ($\pm $ one standard deviation) are shown for the
treated substrate case; they are similar for the other two cases.}
\label{nsides}
\end{figure}

\begin{figure}
\caption{The distribution of crack junction angles for selected runs.
(a) Untreated substrate, no impurities. Solid bars: $d = 0.50 \pm
0.02$ mm, open bars: $d = 3.35 \pm 0.18$ mm. (b) Teflon-coated
substrate, no impurities. Solid bars: $d = 0.80 \pm 0.05$ mm, open
bars: $d = 2.70 \pm 0.08$ mm.}
\label{angles}
\end{figure}


\noindent 

\begin{references}

\bibitem[*]{kelly} present address: Department of Physics, University
of Western Ontario, London ON, Canada.

\bibitem{walker86} J. Walker, Sci. Amer.
{\bf 255}, 204 (Oct. 1986).

\bibitem{basalt} M. P. Ryan and C. G. Sammis,
Geol. Soc. Am. Bull. {\bf 89}, 1295 (1978); J. M. De Graaf and
A. Aydun, Geol. Soc. Am. Bull. {\bf 99}, 605 (1987); M. P. Ryan and
C. G. Sammis, Geol. Soc. Am. Bull. {\bf 89}, 1295 (1978); P.
Budkewitsch and P.-Y. Robin, J. Volcanol. Geotherm. Res.  {\bf 59} 219
(1994).

\bibitem{muller98} G. M\``uller, J. Geophys. Res. {\bf 103B} 15239
(1998).

\bibitem{k17} E. M. Kindle, J. Geol. {\bf 25}, 135 (1917).

\bibitem{t65} J. Q. Tompkins, Geol. Soc.  Am. Bulletin {\bf 76},
1075 (1965); J. T. Neal, {\it ibid.} {\bf 77}, 1327 (1966);
J. Q. Tompkins, {\it ibid.}  {\bf 77}, 1331 (1966).

\bibitem{fracture} L.B. Freund, Dynamic Fracture Mechanics (Cambridge
Univ. Press, New York, 1990).

\bibitem{l62} A. H. Lachenbruch, Geol. Soc. Am. Spec. Pap. {\bf 70} (1962).

\bibitem{alumina} Degussa Canada, Ltd.

\bibitem{al95} C. Allain and L. Limat,  Phys. Rev. Lett. {\bf 74}, 2981 (1995).

\bibitem{sm88} A. T. Skjeltorp and P. Meakin,  Nature (U.K.) {\bf 335}, 424 (1988).

\bibitem{gk94} A. Groisman and E. Kaplan, Europhys. Lett. {\bf 25}, 415 (1994).

\bibitem{wb} J. Webb and T. Beddoe, unpublished.

\bibitem{ppa99} L. Pauchard, F. Parisse, and C. Allain, Phys. Rev. E,
{\bf 59}, 3737 (1999).

\bibitem{kmm98} W. Korneta, S. K. Mendiratta, and J. Menteiro,
Phys. Rev. E {\bf 57}, 3142 (1998).

\bibitem{a94} J. V. Andersen, Phys. Rev. B {\bf 49}, 9981 (1994).

\bibitem{abj94} J. V. Andersen, Y. Brechet, and H. J. Jensen,
Europhys. Lett. {\bf 26}, 13 (1994).

\bibitem{la97} K.-t. Leung and J. V. Andersen, Europhys. Lett. {\bf 38},
589 (1997).

\bibitem{msb93} O. Morgenstern, I. M. Sokolov, and A. Blumen,
Europhys. Lett. {\bf 22}, 487 (1993); J. Phys. A {\bf 26}, 4521
(1993).

\bibitem{smb94} I. M. Sokolov, O. Morgenstern, and A. Blumen,
Macromol. Symp. {\bf 81}, 235 (1994).

\bibitem{hsb96} T. Hornig, I. M. Sokolov, and A. Blumen, Phys. Rev. E
 {\bf 54}, 4293 (1996).

\bibitem{hsb97} U. A. Handge, I. M. Sokolov, and A. Blumen, Europhys. Lett.
{\bf 40}, 275 (1997).

\bibitem{cb97} K. M. Crosby and R. M. Bradley, Phys. Rev. E {\bf 55}
6084 (1997).

\bibitem{k99} S. Kitsunezaki, patt-sol/9905007, unpublished.

\bibitem{griffiths} A. A. Griffiths, Phil. Trans. Roy. Soc. London A
{\bf 221}, 163 (1920).

\bibitem{ys93} A. Yuse and M. Sano, Nature {\bf 362}, 329 (1993);
Physica {\bf D108}, 365 (1998).

\bibitem{ssn94} S.-i. Sasa, K. Sekimoto and H. Nakanishi, Phys. Rev. E {\bf  
50}, R1733 (1994).

\bibitem{ks97} T. S. Komatsu and S.-i. Sasa, Japan. J. Appl. Phys. {\bf 36},  
391 (1997).

\bibitem{hayakawa} Y. Hayakawa, Phys. Rev. E {\bf 49}, R1804 (1994).

\end{references}
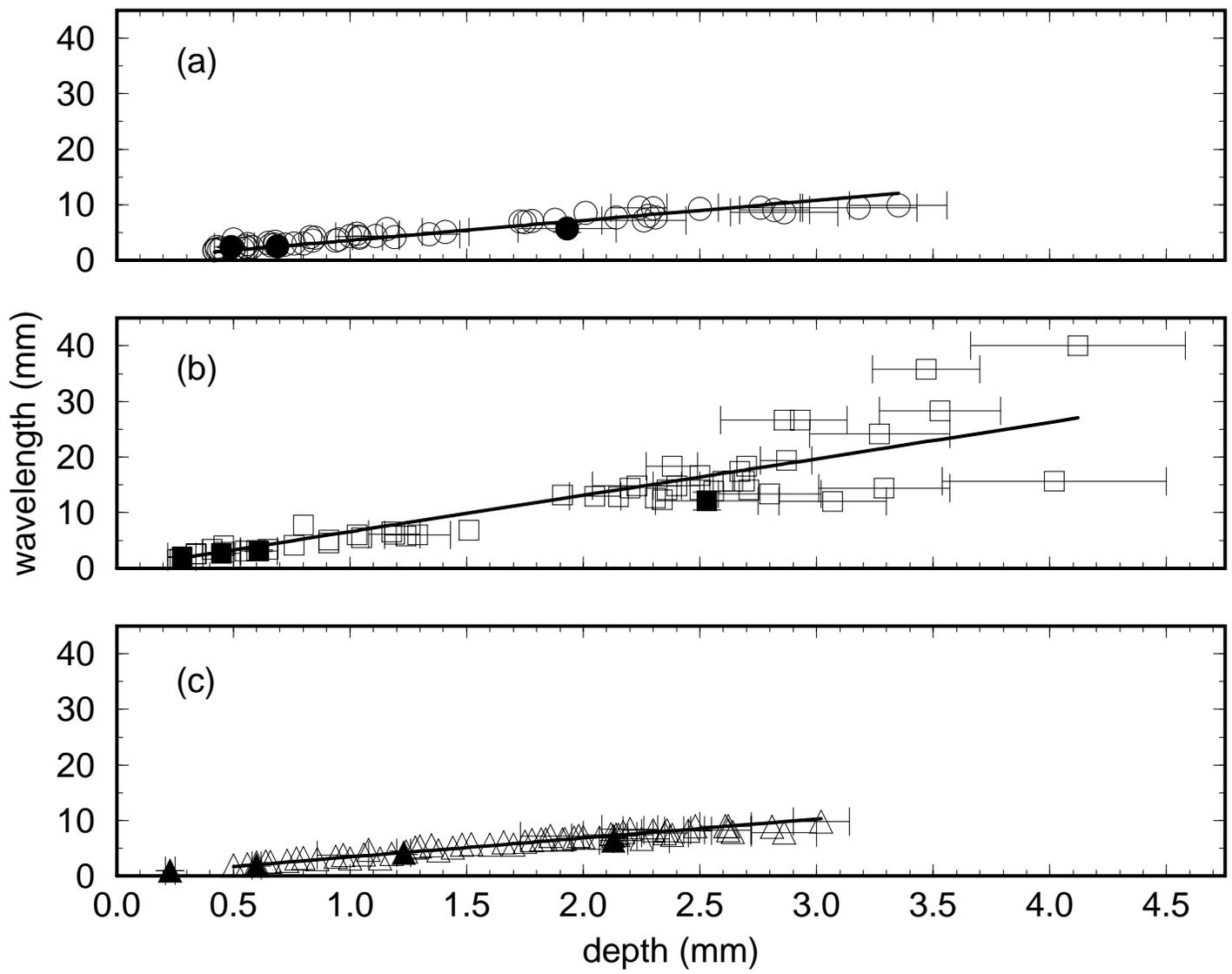
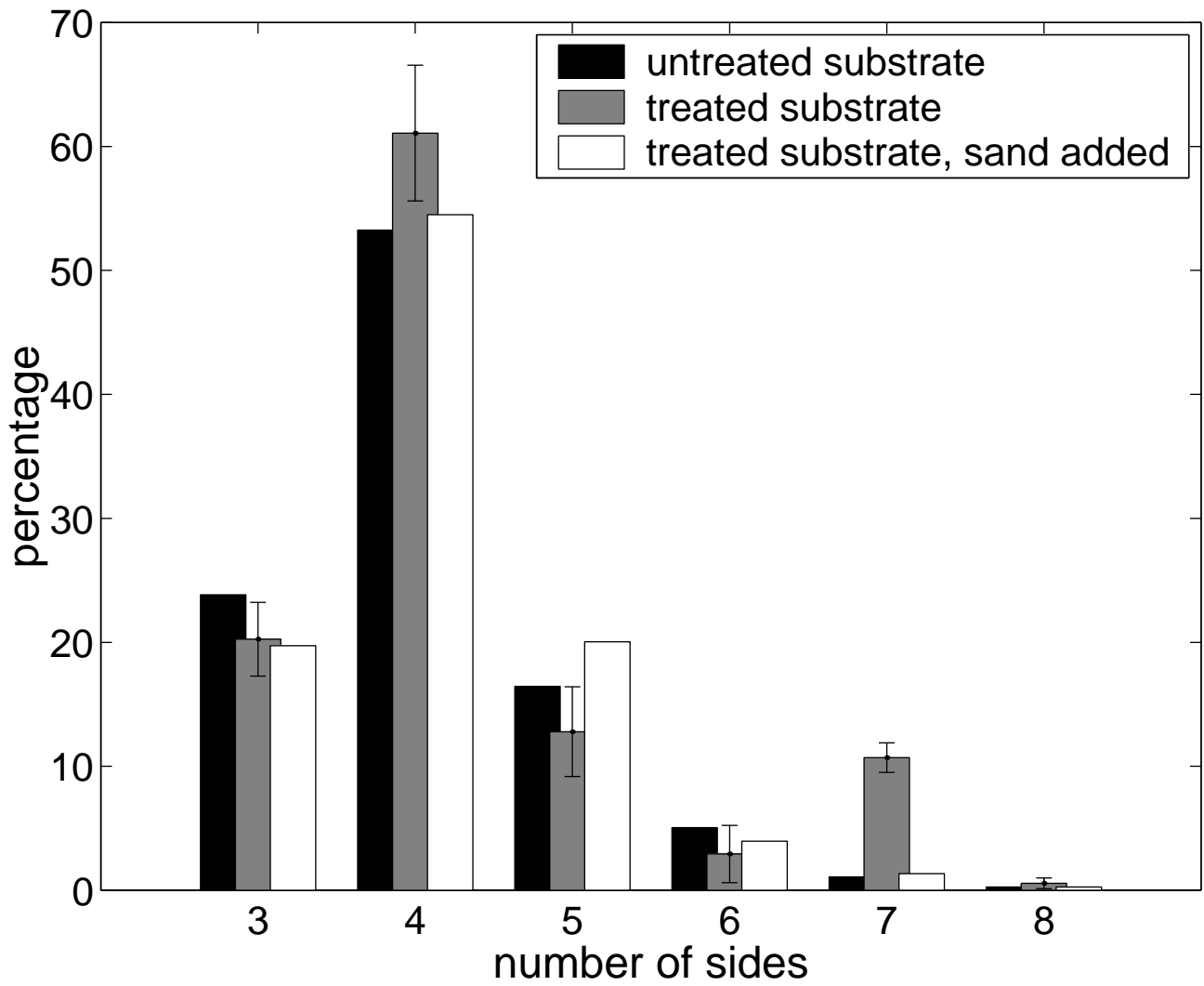
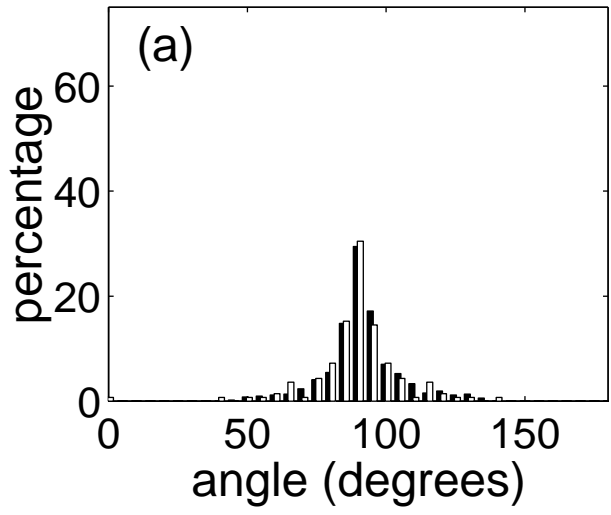
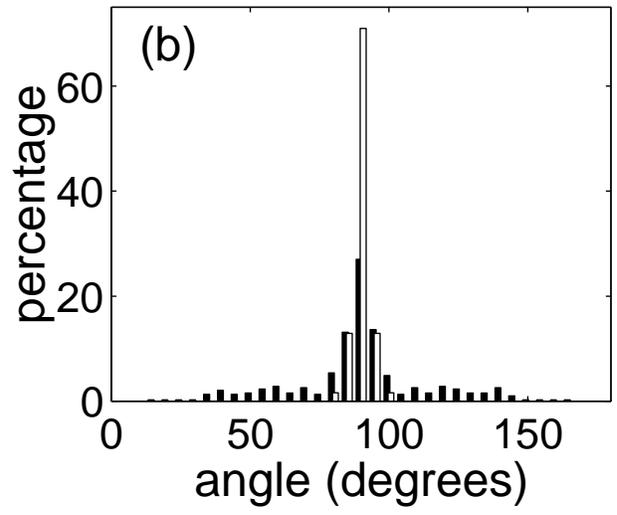
\end{document}